\documentstyle[12pt]{article}
\begin{document}
\def\qb{\overline q}
\def\sb{\overline s}
\def\be{\begin{equation}}
\def\ee{\end{equation}}
\def\bea{\begin{eqnarray}}
\def\eea{\end{eqnarray}}
\begin{center}
\vskip 1.5 truecm
{\bf SOLUTION OF THE OFF-FORWARD LEADING LOGARITHMIC EVOLUTION
EQUATION BASED ON THE GEGENBAUER MOMENTS INVERSION}\\
\vspace{1cm}
A.Shuvaev\\ Theory Department, St.Petersburg
Nuclear Physics Institute\\ 188350, Gatchina, St.Petersburg, Russia.\\
e-mail: {\tt shuvaev@thd.pnpi.spb.ru}
\end{center}
\begin{abstract}
Using the conformal invariance the leading-log evolution of the
off-forward structure function is reduced to the forward evolution
described by the conventional DGLAP equation. The method relies on
the fact that the anomalous dimensions of the Gegenbauer moments of
the off-forward distribution are independent on the asymmetry, or
skewedness, parameter and equal to the DGLAP ones. The integral kernels
relating the forward and off-forward functions with the same Mellin
and Gegenbauer moments are presented for arbitrary asymmetry value.
\end{abstract}
\vspace{1cm}

\noindent
{\bf 1.} The logarithmic evolution of the off-forward structure
function is
described by the generalization of DGLAP equation for the parton
distribution. However the explicit dependence on the longitudinal
momentum transfer makes the splitting functions to be much more
complicated since they include the pieces different in different
kinematics regions \cite{sf}.

From the other hand it is well known that there is the set of
twist-two operators, whose leading-log evolution is exactly diagonal
due to conformal symmetry remaining to be valid at the one loop level
(see e.g. \cite{ohr,BFKL}).
The off-forward matrix elements of these operators are the Gegenbauer
polynomials, they turn into the simple powers of Bjorken $x$ in the
forward kinematics. The Gegenbauer moments of the non-singlet off-forward
function have a simple multiplicative evolution like the forward ones.
There is a mixture of quark and gluon channels for the singlet function
but only between the moments of the same order, the anomalous dimensions
being equal for the forward and off-forward cases. This property
allows, in principle, to reduce the non-diagonal leading-log evolution to
the diagonal one: if one could replace in some way the powers of $x$ in
the solution of the conventional DGLAP equation by the Gegenbauer
polynomials the result returns a solution of the off-forward equation.
However the Gegenbauer polynomials do not form
a complete set on the full interval, where the structure function is
defined, therefore this substitution can not be done straightforward.
It is the aim of the present paper to construct explicitly the
transformation relating the off-forward and forward evolution.

The notations of the paper \cite{BFKL} will be used in what follows
although the final result holds irrespective the particular choice of
the normalization etc. The off-forward structure function can be generally
treated as a function of two variables, $x_1$, $x_2$, playing the role of
the longitudinal momenta of the incoming or outgoing partons (depending
on the sign). The total longitudinal momentum transfer, $\zeta = x_1+x_2$,
defines the asymmetry, or skewedness, of the process, $0\le\zeta\le 1$,
the value $\zeta =0$ describing the forward kinematics.
The Gegenbauer polynomials for the quark ($\qb q$) and gluon ($gg$) channel
read:
$$
R_j^{\qb q}(x_1,x_2)\,=\,R_{n+1}^{\qb q}(x_1,x_2)\,=\,
\sum_{k=0}^n \frac{n!(n+2)!(-1)^k}
{k!(k+1)!(n-k)!(n-k+1)!}\,x_1^k x_2^{n-k}
$$
$$
R_j^{gg}(x_1,x_2)\,=\,R_{n+2}^{gg}(x_1,x_2)\,=\,
\sum_{k=0}^n \frac{n!(n+4)!(-1)^{k+1}}
{k!(k+2)!(n-k)!(n-k+2)!}\,x_1^k x_2^{n-k}.
$$
For a fixed value of $\zeta$ the polynomial moments of the quark and
gluon non-diagonal structure functions
$$
G_{j-1}^{\qb q}(\zeta,t)\,=\,G_n^{\qb q}(\zeta,t)\,=\,
\int dx_1 dx_2\,R_{n+1}^{\qb q}(x_1,x_2)\,
f^{\qb q}(x_1,x_2,t)\,\delta(x_1+x_2 -\zeta)
$$
$$
G_{j-2}^{gg}(\zeta,t)\,=\,G_n^{gg}(\zeta,t)\,=\,
\int dx_1 dx_2\,R_{n+2}^{gg}(x_1,x_2)\,
f^{gg}(x_1,x_2,t)\,\delta(x_1+x_2 -\zeta)
$$
evolve as ($\{r_1,r_2\}=\{\qb q,gg\}$)
\be
\label{Lj}
\frac d{dt} G_j^{r_1 r_2}(\zeta,t)\,=\,\sum_{r_1^\prime r_2^\prime}
\Lambda ^{r_1 r_2}_{r_1^\prime r_2^\prime}(j)\,G_j^{r_1^\prime r_2^\prime}
(\zeta,t),
\ee
$$
t\,=\,\frac 1b\log\biggl(1 + b\frac{g^2}{16\pi^2}\log\frac{Q^2}{\Lambda^2}
\biggr), \quad b=\frac {11}3 N_c -\frac 23 n_f,
$$
where the anomalous dimensions $\Lambda^{r_1 r_2}_{r_1^\prime r_2^\prime}
(j)$ do not depend on $\zeta$ and are determined by the conventional
DGLAP equation. It allows to treat the evolution of the
initial off-forward function $f(x,\zeta-x,t)$ as a product of three
transformations. At first one has to find the function $f_\zeta(x,t)$,
whose Mellin moments are equal (up to normalization) to the Gegenbauer
moments of the initial function $f(x,\zeta-x,t)$ and which includes $\zeta$
only as a parameter. The second step is the forward evolution of this
function governed by the usual DGLAP splitting functions for an appropriate
channel. At the last step one has to evaluate the off-forward function
with Gegenbauer moments equal to the Mellin ones of the diagonal function
evolved at the previous step. This final function provides the solution
of the evolution equation satisfying the initial conditions.

Denoting symbolically the operator of off-forward evolution as
$e^{t h_{ND}}$, where the "hamiltonian" $h_{ND}$ is defined through the
splitting functions, the above transformations can be summarized in the
decomposition
$$
e^{t h_{ND}}\,=\,K_\zeta\,e^{t h_D}\,K_\zeta^{-1}.
$$
Here $e^{t h_D}$ stands for the DGLAP evolution whereas
$t$-independent operator $K_\zeta$ transforms the function with given Mellin
moments to the function with the same Gegenbauer ones.

The auxiliarly function $f_\zeta(x,t)$ evolving according DGLAP equation
is natural to refer as an effective forward, or diagonal, function.

\smallskip
\noindent
{\bf 2.} The problem to find $K_\zeta$ explicitly would be more
or less trivial provided the $R_j$ polynomials form an orthonormal set
on the unit interval, but they are orthogonal only on the smallest
interval $[0,\zeta]$:
\be
\label{normaq}
\int_0^\zeta dx\,x(\zeta-x)\,R_{m+1}^{\qb q}(x,\zeta-x)
R_{n+1}^{\qb q}(x,\zeta-x)\,=\,\zeta^{2n+3}\delta_{mn} /h_{n+1}^{\qb q},
\ee
$$
h_{n+1}^{\qb q}\,=\,\frac{(n+1)(2n+3)}{n+2},
$$
\be
\label{normag}
\int_0^\zeta dx\,x^2(\zeta-x)^2\,R_{m+2}^{gg}(x,\zeta-x)
R_{n+2}^{gg}(x,\zeta-x)\,=\,\zeta^{2n+5}\delta_{mn} /h_{n+2}^{gg},
\ee
$$
h_{n+2}^{gg}\,=\,\frac{(n+2)(n+1)(2n+5)}{(n+3)(n+4)}.
$$

To clarify the idea we start from a simple example --
how to restore the function knowing its usual moments,
\be
\label{Mn}
M_n\,=\,\int dx\,x^n\,f(x)
\ee
only for integer $n$ without making use of the Mellin transformation
based on the continuation to the complex indices. Formally the solution
can be written as
\be
\label{fx}
f(x)\,=\,\sum_{n=0}^\infty M_n\, \delta^{(n)}(x) \,\frac 1{n!}(-1)^n,
\ee
where $\delta^{(n)}(x)$ is the $n$-th derivative of the
$\delta$-function. To make a convergent solution we rewrite the
$\delta$-functions as
\be
\label{disc} \delta^{(n)}(x)\,=\,-\frac
1{2\pi i}(-1)^n n!\left[\frac 1{(x+i\epsilon)^{n+1}}\,-\, \frac
1{(x-i\epsilon)^{n+1}}\right]
\ee
and define on the complex plane the function
\be
\label{Fz}
F(z)\,=\,\sum_{n=0}^\infty \frac{M_n}{z^{n+1}}.
\ee
The imaginary part of it, or discontinuity at the real axis,
\be
\label{Im}
\Im F(x)\,=\,\frac 1{2i}\,\bigl[F(x+i\delta)\,-\,F(x-i\delta)\bigr],
\ee
gives the function $f(x)=-1/\pi \Im F(x)$. If the moments $M_n$ grow
not very fast with $n$,
which is really needed for the Mellin transform to be localized at
$x \le 1$, the function $F(z)$ is analytical for $|z| \ge 1$ so that
$f(x)\not =0$ only for $|x| \le 1$.

Actually this derivation is nothing more than the usual procedure
relating the matrix elements of the local operators to the Mellin
moments of the forward structure function. Note that any finite
number of the terms in eq.(\ref{fx}) has only the single point $x = 0$ as
a support whereas the support of the infinite sum spreads
over the whole interval, where $f(x) \not =0$, that is over the integration
region for the moments (\ref{Mn}). Note also that this interval can generally
include both the positive and negative $x$ values.

Now we apply the same trick to find the kernel $K_\zeta(x,y)$
relating the off-forward and effective forward functions in the
quark channel,
\be
\label{Kqq}
f(x,\zeta)\,=\,\int dy K_\zeta(x,y)\,f(y).
\ee
We use $f(x,\zeta)$ and $f(x)$ as short-hands for $f^{\qb q}(x,\zeta-x,t)$
and $f_\zeta^{\qb q}(x,t)$.
First of all we formally express the function $f(x,\zeta)$ through
its Gegenbauer moments,
\be
\label{Gn}
G_n^{\qb q}(\zeta)\,=\,\int dx\,R_{n+1}^{\qb q}(x,\zeta-x)\,f(x,\zeta),
\ee
as
\be
\label{fxz}
f(x,\zeta)\,=\,\sum_n \int_0^\zeta du \bigl[u(\zeta-u)\bigr]^{n+1}
\delta^{(n)}(x-u)\,G_n^{\qb q}(\zeta)\,\frac{(n+2)!}{\bigl[(n+1)!\bigr]^2}\,
\frac{h_{n+1}^{\qb q}}{\zeta^{2n+3}}.
\ee
This formula can be easily checked by taking into account the
condition (\ref{normaq}) and that
$$
R_{n+1}^{\qb q}(x,\zeta-x)\,=\,\frac{(n+2)!}{\bigl[(n+1)!\bigr]^2}\,\frac
1{x(\zeta-x)}\,\frac{\partial^n}{\partial x^n}\,
\bigl[x(\zeta-x)\bigr]^{n+1}.
$$

Really we are interested in the case when the Gegenbauer moments in
the RHS of (\ref{fxz}) are proportional to the Mellin moments of a function
$f(x)$. Since
$$
R_{n+1}^{\qb q}(x,-x)\,=\,(-1)^n\,\frac{(2n+2)!}{\bigl[(n+1)!\bigr]^2}\,x^n
$$
the normalization is natural to choose as
$$
G_n^{\qb q}(\zeta)\,=\,(-1)^n \frac{(2n+2)!}{\bigl[(n+1)!\bigr]^2}\,M_n.
$$
It ensures $K_{\zeta=0}(x,y)=\delta(x-y)$.
Then the sum (\ref{fxz}) takes the form
\be
f(x,\zeta)\,=\,\sum_{n=0}^\infty \int_0^\zeta du
\bigl[u(\zeta-u)\bigr]^{n+1}
\delta^{(n)}(x-u)\,(-1)^n\,\frac{(2n+3)!}{n!\bigl[(n+1)!\bigr]^2}\,
\zeta^{-2n-3}\,M_n.
\ee

Replacing $\delta$-functions by the discontinuity (\ref{disc})
and changing the variables $u=\zeta s$, we again express the solution
through the imaginary part:
\bea
f(x,\zeta)\,&=&\,-\frac 1\pi\,\Im F_q(z,\zeta)  \nonumber \\
F_q(z,\zeta)\,&=&\,\int_0^1 ds \,\sum_{n=0}^\infty \left[
\frac{s(1-s)}{z-\zeta s}\right]^{n+1}\frac{(2n+3)!}{\bigl[(n+1)!\bigr]^2}
\,M_n. \nonumber
\eea
Substituting here the Mellin moments (\ref{Mn}) we arrive at the
expression for the integral kernel (\ref{Kqq}):
\be
\label{ker}
K_\zeta(x,y)\,=\,-\frac 1{2\pi}\,\Im \int_0^1 ds \,\sum_{n=0}^\infty
\frac{(2n+3)!}{\bigl[(n+1)!\bigr]^2}\left[
\frac{s(1-s)}{x-\zeta s}y\right]^{n+1}\frac 1y.
\ee

It is easily to check that
$$
K_{\zeta=0}(x,y)\,=\,-\frac 1\pi \Im \sum_{n=0}^\infty
\left(\frac xy \right)^{n+1}\frac 1y\,=\,-\frac 1\pi \Im \frac 1{x-y}\,
=\,\delta(x-y),
$$
where the imaginary part is understood as in eq.(\ref{Im}).

Similarly to the Mellin transform case the support in the infinite sum
in (\ref{fxz}) is different from those for any finite number of the
terms which is actually $[0,\zeta]$ interval. Note that the integration
in the RHS of eq.(\ref{Kqq}) covers both the positive and negative $y$ value,
the effective diagonal function $f(y)$ being the quark structure function
for the positive argument and minus antiquark function for the negative $y$.

The sum in the kernel (\ref{ker}) can be calculated in the closed form.
Indeed, using the identity
$$
\frac{2n!}{\bigl[n!\bigr]^2}\,=\,\frac 1\pi
\,4^n\,\frac{\Gamma(n+\frac 12)\Gamma(\frac 12)}{\Gamma(n+1)} \,=\,
\frac 1\pi
\,4^n \int_0^1 dv\, v^{n-\frac 12}(1-v)^{-\frac 12}
$$
it can be rewritten as
$$
\sum_{n=0}^\infty \frac{(2n+3)!}{\bigl[(n+1)!\bigr]^2}\,z^{n+1}\,=\,
\bigl(2z\,\frac \partial{\partial z}\,+\,1\bigr)\,
\sum_{n=1}^\infty \frac{2n!}{\bigl[n!\bigr]^2}\,z^n\,=
$$
$$
=\,\frac 1\pi \,\bigl(2z\,\frac \partial{\partial z}\,+\,1\bigr)\,
\int_0^1 dv\, v^{-\frac 12}(1-v)^{-\frac 12}\,\frac{4vz}{1-4vz}.
$$
Taking the imaginary part one gets
\bea
K_\zeta(x,y)\,&=&\,\int_0^1 ds\,\frac 1 \pi\,\int_0^1 dv \,
v^{-\frac 12}\,(1-v)^{-\frac 12}\,4v\, s\sb\, y \times  \nonumber \\
&\times&\,\bigl[\,3\,\delta(x-\zeta s -4v \,s \sb \,y)\,
-\,8 \,v \,s\sb y\,\delta^\prime(x-\zeta s -4v \,s \sb \,y\bigr],
\nonumber
\eea
$$
\sb\,\equiv\,1\,-\,s,
$$
or, finally
\be
\label{Kq}
K_\zeta(x,y)\,=\,-\frac 1 \pi \frac 1{|y|}\,\int_0^1 ds \,
\left[\frac{z(s)}{1-z(s)} \right]^{\frac 32} \,\theta\bigl(z(s)\bigr)
\theta\bigl(1-z(s)\bigr),
\ee
\be
\label{zs}
z(s)\,=\,\frac{x-\zeta s}{4 s \sb\,y}.
\ee
The possible singularity occurring for $z(s)\to 1$ should be treated
in the principle value sense.

\smallskip
\noindent
{\bf 3.} Consider now the inverse kernel allowing to construct
the function, which Mellin moments are equal to the Gegenbauer ones,
that is in the normalization we choose
$$
M_n(\zeta)\,=\,(-1)^n \frac{\bigl[(n+1)!\bigr]^2}{(2n+2)!}\,G_n^{\qb q}(\zeta).
$$
Since
$$
G_n^{\qb q}(\zeta)\,=\,(-1)^n \frac{(n+2)!}{\bigl[(n+1)!\bigr]^2}\,\int dy\,
f(y,\zeta)\,\frac 1{y(\zeta - y)}\,\frac{\partial^n}{\partial y^n}
\bigl[\,y(\zeta-y)\bigr]^{n+1}
$$
the Mellin moments are
\be
M_n(\zeta)\,=\,(-1)^n \int dy\,f(y,\zeta)\,\frac 1{y(\zeta - y)}\,
\frac{(n+2)!}{(2n+2)!}\,
\frac{\partial^n}{\partial y^n} \bigl[\,y(\zeta-y)\bigr]^{n+1}.
\ee
Recalling the formula (\ref{Fz}) the kernel is given by the imaginary part
$$
K^{(-1)}(x,y)\,=\,-\frac 1\pi\,\frac 1{y(\zeta - y)}\,\Im q(x,y)
$$
of the sum
\be
\label{qzy}
q(z,y)\,=\,\int_0^1 ds\,\sum_{n=0}^\infty \frac {(-1)^n}{z^{n+1}}
\,\frac{2n+3}{n!}\,s^{n+2}\, \sb^n\,
\frac{\partial^n}{\partial y^n} \bigl[-s\sb\,y(\zeta-y)\bigr]^{n+1}.
\ee
Again, this sum can be converted into the compact expression. To this
end we rewrite it as
$$
q(z,y)\,=\,\int_0^1 ds\,\frac s\sb \,z\frac \partial{\partial z}
\biggl(1\,-\,2z\frac \partial{\partial z}\biggr)\,p(z,y),
$$
where
$$
p\,=\,y\,+\,\sum_{n=1}^\infty \frac 1{z^n}\,\frac 1{n!}\,
\frac{\partial^{n-1}}{\partial y^{n-1}}
\bigl[-s \sb\,y(\zeta-y)\bigr]^n.
$$

The series for $p$ is actually the Lagrange's Expansion \cite{AS}
for the solution of the equation
$$
p\,=\,y\,-\,\frac 1z\,s \sb\,p\,(\zeta-p)
$$
regular at $z=\infty$. Thus
\be
p(z,y)\,=\,\frac 12 \zeta\,+\,\frac z{2\,s\sb}\,-\,\frac 1{2\,s\sb}\,
\bigl[z^2+2 s\sb (\zeta-2y)z + (s\sb \,\zeta)^2 \bigr]^{\frac 12}.
\ee

Only the square root term survives in the imaginary part.
Taking the derivatives one gets the following kernel
\be
\label{Kqi}
K^{(-1)}_\zeta(x,y)\,=\,\frac 1{2\pi}\,\frac x{y(\zeta - y)}\,
\int_0^1 \frac {ds}{\sb^2}\,Q_q(x,y)\,R^{-\frac 32}(x,y)
\theta \bigl(R(x,y)\bigr),
\ee
\bea
\label{Rxy}
R(x,y)\,&=&\,2\,(2y-\zeta) x s\sb \,-\,x^2\,-\,(x s\sb \zeta)^2\\
Q_q(x,y)\,&=&\,\bigl(3x^2+(s\sb \zeta)^2\bigr) (2y-\zeta)s\sb - x^3 -
3x (s\sb \zeta)^2. \nonumber
\eea
The variable $x$ takes here the positive and negative values, and
the kernel $K^{(-1)}_\zeta$ produces both the quark and (minus) antiquark
effective diagonal function.

\smallskip
\noindent
{\bf 4.} Now we turn to the gluon channel. Taking into account the
normalization of the gluon polynomials (\ref{normag}) and that
$$
R_{n+2}^{gg}(x,\zeta-x)\,=\,-\frac{(n+4)!}{\bigl[(n+2)!\bigr]^2}\,\frac
1{x^2(\zeta-x)^2}\,\frac{\partial^n}{\partial x^n}\,
\bigl[x(\zeta-x)\bigr]^{n+2}
$$
the function $f(x,\zeta)$ having Gegenbauer moments $G_n^{gg}(\zeta)$
formally is
\be
f(x,\zeta)\,=\,-\sum_{n=0}^\infty \int_0^\zeta du
\bigl[u(\zeta-u)\bigr]^{n+2}
\delta^{(n)}(x-u)\,\frac{2n+5}{n!}\,G_n^{gg}(\zeta)\,\zeta^{-2n-5}.
\ee
The expansion
$$
F_g(z,\zeta)\,=\,\frac 1y\,\int_0^1 ds\,s\sb \sum_{n=0}^\infty
\left(\frac{s\sb y}{z-s\zeta}\right)^{n+1}\frac{(2n+5)!}{\bigl[(n+2)!\bigr]^2}
$$
determines the kernel,
$$
{\overline K}_\zeta(x,y)\,=\,-\frac 1\pi\,\Im F_g(z,\zeta),
$$
relating the functions, whose Gegenbauer and Mellin moments are
proportional,
$$
G_n^{gg}(\zeta)\,=\,(-1)^{n+1}\frac{(2n+4)!}{\bigl[(n+2)!\bigr]^2}\,M_n.
$$
As well as for the quark case the coefficient is fixed here by
$R_{n+2}^{gg}(x,-x)$ value and ensures ${\overline K}_{\zeta=0}(x,y)=
\delta(x-y)$. After rewriting it as
\bea
F_g(z,\zeta)\,&=&\,\frac 1y\,\int_0^1 ds\,s\sb \,\frac{z-s\zeta}{s\sb y}\,
\left[-6\frac{s\sb y}{z-s\zeta}\right.\,+\nonumber \\
&+&\left.\,\sum_{n=0}^\infty\left(\frac{s\sb y}{z-s\zeta}\right)^{n+1}
\frac{(2n+3)!}{\bigl[(n+1)!\bigr]^2}\right]\,= \\
&=&\,-\frac 1y\,+\,\frac 1{y^2}\int ds\,\bigl(z-s\zeta\bigr)
\sum_{n=0}^\infty\left[\frac{s (1-s)}{z-s\zeta}y\right]^{n+1}
\frac{(2n+3)!}{\bigl[(n+1)!\bigr]^2} \nonumber
\eea
it expresses in terms of the sum calculated before, eq.(\ref{ker}). Thus
\be
\label{Kg}
{\overline K}_\zeta(x,y)\,=\,-\frac 1\pi\,\frac 1{y|y|}\,\int ds\,
\bigl(z-s\zeta\bigr)
\left[\frac{z(s)}{1-z(s)} \right]^{\frac 32} \,\theta\bigl(z(s)\bigr)
\theta\bigl(1-z(s)\bigr),
\ee
where $z(s)$ is the same (\ref{zs}). This kernel like the quark one implies
both signs of $y$, the negative $y$ stands for the antigluon structure
function.

The derivation of the inverse gluon kernel repeats those for the quark
channel. The kernel is given by the imaginary part of the function $g(x,y)$,
$$
{\overline K}_\zeta^{(-1)}(x,y)\,=\,-\frac 1\pi \,\frac 1{y^2(\zeta-y)^2}\,
\Im g(x,y),
$$
$$
g(z,y)\,=\,\sum_{n=0}^\infty \frac{(-1)^n}{z^{n+1}}\,\frac{(n+4)!}{(2n+4)!}
\frac{\partial^n}{\partial y^n}\,\bigl[y(\zeta-y)\bigr]^{n+2}.
$$
Comparing this expression with the quark function $q(z,y)$ (\ref{qzy})
it is easy to check that
$$
\frac \partial{\partial y}g(z,y)\,=\,3y(\zeta-y)\,+\,z\,\biggl(
z\frac \partial{\partial z}\,-\,2\biggr)\,q(z,y)
$$
$$
g(z,y=0)\,=\,0.
$$
The integration yields

\be
\label{Kgi}
{\overline K}_\zeta^{(-1)}(x,y)\,=\,\frac 1{4\pi} \,\frac 1{y^2(\zeta-y)^2}\,
\int_0^1\frac {ds}{\sb^2}\,Q_g(x,y)\,R^{-\frac 32}(x,y)
\theta \bigl(R(x,y)\bigr),
\ee
where the polynom $R(x,y)$ is given by eq.(\ref{Rxy}) and
\bea
Q_g(x,y)\,&=&\,\bigl(x^2+3\zeta(s\sb)^2\bigr)^2(2y-\zeta)x - 10\bigl(
2y^2-2y\zeta+\zeta^2\bigr)x^2\zeta^2(s\sb)^3 \nonumber\\
&-& 3\bigr(2y-\zeta\bigr)^2 x^4 s\sb - 3\zeta^6(s\sb)^5.
\eea

\smallskip
\noindent
{\bf 5.} The expressions (\ref{Kq}), (\ref{Kqi}), (\ref{Kg}) and (\ref{Kgi})
are the main results of this paper. They completely reduce the off-forward
leading-log evolution to the well known DGLAP equation as it has been outlined
above. To carry through this transformation one has to specify the functions
$f(x_1,x_2)$. Taking for definiteness X.Ji notations one has for the quark
channel
$$
f^{\qb q}(x_1,x_2)\,=\,\bigl(1\,+\,\frac 12 \xi \bigr)\,H_q(x,\xi),
$$
$$
x\,=\,\frac{(x_1+x_2)/2}{1-(x_1-x_2)/2},\quad
\xi\,=\,\frac{(x_1-x_2)/2}{1-(x_1-x_2)/2}
$$
and the same relation for the function $E$. It is worth to emphasize that
the normalization as well as the spinor pre-factors appearing in
the definition of the structure functions are unessential from the evolution
viewpoint.

The relation for the gluon channel includes additional $x$:
$$
f^{gg}(x_1,x_2)\,=\,\bigl(1\,+\,\frac 12 \xi \bigr)\,x\,H_g(x,\xi).
$$

The kernels $K_\zeta^{(-1)}$ convert the initial off-forward functions
to the effective forward ones $q_\zeta(x,t_0)$ and $x g_\zeta(x,t_0)$
parametrically dependent on $\zeta$. After the functions $q_\zeta(x,t)$
and $g_\zeta(x,t)$ have evolved according to DGLAP the kernels $K_\zeta$ turn
the results, $g_\zeta(x,t)$ and $x g_\zeta(x,t)$, into the final off-forward
functions $H_q(x,\xi)$ and $x H_g(x,\xi)$. The reason for the extra $x$
in front of the gluon function, $x g_\zeta$, is that the anomalous dimensions
$\Lambda ^{r_1 r_2}_{r_1^\prime r_2^\prime}(j)$ (\ref{Lj}) taken from the
DGLAP are associated with the Gegenbauer polynomials of the order $j-1$
for the quarks and $j-2$ for the gluons channels.

\vspace{0.5cm}
\noindent
The author is grateful to M.G.Ryskin
for helpful discussions.

\vspace{0.5cm}
\noindent
This work is financially supported by RFFI grant 98-02-17629.

\end{document}